\documentclass[%
 reprint,
 amsmath,amssymb,
 aps,
pre]{revtex4-1}

\usepackage{todonotes}
\usepackage{units}

\usepackage{graphicx}
\usepackage{dcolumn}
\usepackage{bm}

\usepackage{tikz}
\usetikzlibrary{angles,quotes,calc}
\usepackage{fancyref}

\pgfmathsetmacro\gratio{1.04839}

\newcommand{\Km}{K_{\max}}
\newcommand{\Q}{\mathbb{Q}}

\begin{document}

\preprint{APS/123-QED}

\title{Ergodicity in Golden Billiards}

\author{Joseph Seaward}
\affiliation{Universit\'e Paris 13, Sorbonne Paris Cit\'e, Laboratoire de physique des lasers, F-93430, Villetaneuse, France
}
 \affiliation{CNRS, UMR7538, F-93430, Villetaneuse, France}
 \email{joseph.seaward@univ-paris13.fr}
%


\date{\today}

\begin{abstract}
    This comment is an analysis of the results presented by Wang et al. in their their 2014 paper on irrational right triangular billiards. They submit numerical evidence that these billiards are a novel kind of nonergodic, incompatible with KAM theory, at least in the ``strongly irrational'' case, typified by one of the angles being defined by the golden ratio. We offer an explanation of their results as well as a discussion of ergodicity. We suggest that the system is likely to be ergodic and offer a way to reconcile it with KAM theory if it is not.
\end{abstract}


\maketitle

\subsection{Introduction}
\label{sec:intro}

    In their paper\cite{wang2014nonergodicity}, Wang et al. take the problem of two masses moving without friction between two walls and convert it into a billiard problem of a virtual particle moving inside a right triangle, of angle $\alpha/2$ defined such that $\cos{\alpha} = \frac{m_1-m_2}{m_1+m_2}$, without ever hitting a corner. They then reparameterize the problem in such a way that the angle of the velocity of a virtual particle in the triangle is just $\theta = \varphi + K\alpha$ with $\varphi \in \{\theta_0, -\theta_0, \pi - \theta_0, \pi + \theta_0\}$, $\theta_0$ is the initial angle of the velocity, taken to be an irrational multiple of $\pi$, and $K \in \mathbb{Z}$. The angle $\varphi$ holds the information about the initial angle and the quadrant in which the angle of the final velocity lies. So we get four sets of $K\alpha$, one for each value of $\varphi$. The rest of the information is contained in $K$ which evolves by only two rules depending on whether the virtual particle hits the hypotenuse of the triangle (encoding a particle-particle collision) or one of the legs (encoding a particle-wall collision). They are
\begin{equation}
    \begin{split}
      K' = K + 1 \quad &\mathrm{(hypotenuse/particle\: collision)} \\
      K' = -K \quad &\mathrm{(leg/wall\: collision.)}
    \label{eq:K rules}
\end{split}
\end{equation}

    They then run numerical simulations with $\alpha = \pi M/N$ looking at the value of $K$ as well as various invariants of the system (e.g. $\tau_K$, $\tau*$) that we will address later. The principal surprise of their results is that in the ``strongly irrational'' case, $M,N : \lim_{N\rightarrow \infty} M/N = \phi$ where $\phi$ is the golden ratio of $(\sqrt{5} - 1)/2$, the value of $K$ is localized around its initial value, $K_0$. That is to say it appears nonergodic as it seems to retain memory of its initial condition. $K$ has non-zero probability of reaching any value in its space of available states but the frequency of its visits falls off exponentially as one moves away from the initial value. If $\alpha$ is a rational multiple of $\pi$ (i.e. $\alpha \in \mathbb{Q}\pi$), $|K|\rightarrow \pm \infty$ with time, as is expected since billiards of all rational and near-rational polygons are known to be ergodic\cite{Vorobets_1996}.
     
    The authors posit that, in the strongly irrational case, the system is nonergodic with respect to its velocity angles, $\theta$; as in principle they can be anything but there is a vanishing probability that they will take values corresponding to very large $|K|$. They go on to suggest that this is a new kind of nonergodicity as it does not correspond to a phase space neatly divided into regions of positive measure in accordance with KAM theory\cite{lichtenberg1983regular}.

\subsection{Explanation of Numerical Results}
    Looking at the strongly irrational case of $\alpha = \pi\phi$ it is helpful to remind ourselves what the authors are doing. They take th fraction, $M/N$, such that $M/N \rightarrow \phi$ as $N \rightarrow \infty$. This makes $M$ and $N$ Fibonacci numbers of succeeding indices. They then simulate a system with $\alpha = \pi \frac{M}{N}$ and look at various scales related to how long it takes the orbit to close as they increase $N$. For the case of average time to closure, $\left<\tau_K \right>$, this is equivalent to asking how the orbit of the approximant, which is of size $2N_n$, grow with better approximations: that is, with growing $n$. From Binet's formula for the $n^{th}$ Fibonacci number, $F_n$, written 
\begin{equation}
    F_n = \frac{\phi^{-n} - (-\phi)^n}{2\phi+1}
    \label{eq:Binet}
\end{equation}    
    and recalling that $N_n = F_{n+1}$ it can be seen that $N_n \sim \phi^n$, or changing bases of the exponential, $e^{n\ln{\phi}}$. The value $\ln{\phi} = 0.481$ being in close agreement with their fit of $e^{0.49n}$.
    
    They define $\tau^*(N,N')$ as the time until two subsequent rational approximants of $\phi$ give sequences of identical bounces for the same initial angle and thus $\pi \frac{M}{N}$ is equivalent to $\phi$ up to $\tau^*(N,N')$, which should get longer as the index of $N$ increases. They give the dependence as $\tau^* \sim N$. This is not surprising since, if one makes the denominator of the approximant twice as large and thus makes it twice as accurate, the trajectory governed by it should match for twice as long.
    
    They also plot $n^*_K$ which they define as the number of $K$ values visited at time $\tau^*$. This is asking how fast the maximum value of $K$, $K_{\max}$, grows with $N$ which is to say, how fast it grows with time for $\alpha = \phi\pi$. The question of how $K_{\max}$ increases is a question of the rate of double-leg hits. Imagine $K$ is at some arbitrary positive value less than $K_{\max}$. After a $K \rightarrow K+1$ hypotenuse collision, the virtual particle must strike a leg and $K\rightarrow-K$. It can then either strike the hypotenuse again, increasing by one and decreasing its absolute value, or it may hit the other leg taking it back to its previous positive value. The next collision must necessarily be with the hypotenuse and increase the absolute value. These double-leg hits are the only way $|K|$ can grow as if this never happened it would go: $K = 0 \rightarrow 1 \rightarrow -1 \rightarrow 0$ forever. This would correspond to $\alpha = \nicefrac{\pi}{4}$ and $\theta_0$ being on the horizontal or vertical axis. It has already been shown that almost all trajectories starting perpendicular to one of the legs of the triangle form closed orbits\cite{cipra1995periodic}. This is avoided by Wang et al. even in the rational case by insisting $\theta_0 \notin \mathbb{Q}\pi$. However, for $n^*_K$ to grow subsequently, the leg-leg hits must occur when $K = K_{\max}$, so it is little wonder that $n^*_K$ grows slowly.

    If one ignores the details of the dynamics and squints enough, we can try to treat the long-term behavior probabilistically. After hitting the hypotenuse and then a leg, let the probability that the particle will hit the other leg be $p$. It then must hit the hypotenuse and then another leg. Assuming that leg vs. hypotenuse hits are history-independent (which they in fact are not, as attested to by the jumps in $n^*_K$ if nothing else) the probability of two leg-leg hits in a row will be $p^2$. Since $K = \Delta K + K_0$ is localized around its initial value, $K_0$, on average, $K$ will be $K_0$. If we let $K_0=0$ for convenience it will take on average $K_{\max} + 1$ sequential leg-leg hits to raise the value of $K_{\max}$. Assuming that these leg-leg hits happen at some constant average rate in time, we can say that the probability of getting $k$ double leg hits in a row in time $t$ is $p_k(t) = p^kt$. If one fixes $k$ (and $K_{\max}$ will be fixed until this string of $k>K_{\max}$ hits occurs) and takes the long-time limit, $p_k\rightarrow1$. Now since the meaning of ``$\sim$'' is ``ratio approaches one'' we can say that in the long time limit $t\sim p^{-k}$. With some rearranging and by recalling that $t\sim\tau^*\sim N$ and that $k\sim K_{\max} \sim n^*_K$ we may write in their notation
    \begin{equation*}
        n^*_K \sim \log_{\frac{1}{p}}N \,.
        \label{eq:kgoesas}
    \end{equation*}
    This will agree with their results if $1/p \approx e$.
    
    \begin{figure}
        \centering
            \begin{tikzpicture}[scale = 1.3, every node/.style={scale=1.3}]
            
            \coordinate (O) at (0,0);
            \coordinate (X) at (5,0);
            \coordinate (x) at (2.5,0);
            \coordinate (Y) at (5,5.2419478);
            \coordinate (y) at (5,4.2);
            
            \draw[thick] (O) -- (X) -- (Y) --  (O);
            \draw[ultra thin] (X) rectangle +(-.3,.3);
            
            \draw[loosely dashed] (x) -- (Y);
            \filldraw[black] (x) circle (2pt) node[below, inner sep = 6pt]{$x$};
            
            \draw[loosely dashed] (y) -- (O);
            \filldraw[black] (y) circle (2pt) node[right, inner sep = 6pt]{$y$};
            
            \pic[draw, - , thin, "$\theta_x$", angle eccentricity =1.4, angle radius = .7cm]{angle = X--x--Y};
            \pic[draw, - , thin, angle eccentricity =1.4, angle radius = .62cm]{angle = X--x--Y};
            
            \pic[draw, - , thin, "$\theta_y$", angle eccentricity =1.4, angle radius = .75cm]{angle = O--y--X};
            \pic[draw, - , thin, angle eccentricity =1.4, angle radius = .68cm]{angle = O--y--X};
            \pic[draw, - , thin, angle eccentricity =1.4, angle radius = .61cm]{angle = O--y--X};
            
            \pic[draw, - , thick, "{$\alpha/2$}",angle eccentricity =-.7, angle radius = .65cm]{angle = X--O--Y};
        
            \end{tikzpicture}
        \caption{Geometrically estimating the probability of $K$ increasing by calculating angles subtended by opposite legs (see text as to why this increases $K$.) A particle colliding at $x$ ($y$) sees the other leg subtended by angle $\theta_x(x)$ ($\theta_y(y)$). The angle $\alpha/2$ is as defined in \textsection\ref{sec:intro}.}
        \label{fig:triangle}
    \end{figure}
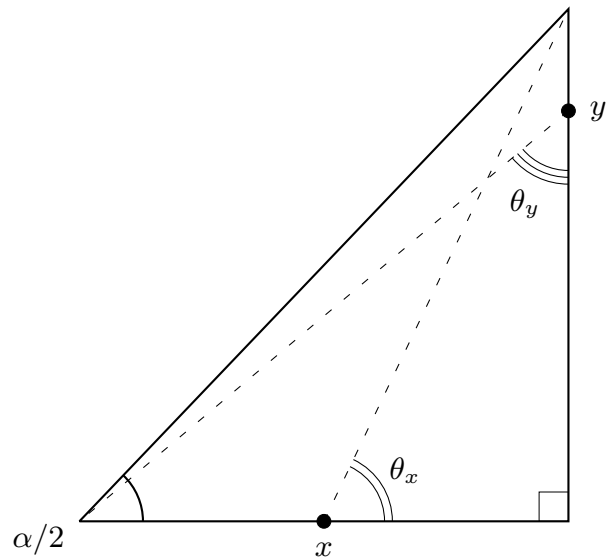
    
    Then let us try and approximate $p$. Let us assume that the particle is equally likely to hit any spot on the leg, and is equally likely to go in any direction. Under this assumption the probability that the particle will hit one leg after the other will be equal to the angle subtended by the other leg from a point on the leg it just hit divided by $\pi$, the total angle it will scatter into. Let the hypotenuse be of unit length and let $\theta_y$ be the angle subtended by the other leg when the particle is on the leg opposite the angle equal to $\alpha/2$ and vice-versa for $\theta_x$ (see \fref{fig:triangle}.) Some geometry then gives
    \begin{equation}
    \begin{split}
        \theta_x(x) = \tan^{-1}\left( \frac{\sin{(\alpha/2)}}{\cos{(\alpha/2)} - x}\right) \\ 
        \theta_y(y) = \tan^{-1}\left( \frac{\cos{(\alpha/2)}}{\sin{(\alpha/2)} - y}\right) \,.
        \label{eq:thetas}
    \end{split}
    \end{equation}
    Averaging the average of \fref{eq:thetas} together should give an estimate of $p$, $\left<p\right> = \frac{\left<\theta_x\right>+\left<\theta_y\right>}{2\pi}$. Calculating the reciprocal of this numerically gives $\left<p\right>^{-1} = 2.8011...$, only a 3\% difference from $e=2.7183...$. This is not a minimum difference from $e$\footnote{This occurs at $\alpha/2 =  \nicefrac{\pi}{4}$ and gives $\left<p\right>^{-1} = 2.7741...$ a difference of 2\%. This also constitutes a maximum for $\left<p\right>$} but would be difficult to distinguish numerically. If this analysis holds, then $n^*_K$ should grow most slowly for $\alpha$ an irrational angle close to $\pi/2$, even if it is not as far from rationality as $\phi$ is.

\subsection{The Question of Ergodicity}

	$K$ forms an invariant measure of the phase space of $\theta$ since the action of \eqref{eq:K rules} clearly cannot take it outside of the set of angles measured by $K$. Wang et al. argue that it should not be full measure since it is exponentially localized and therefore there is a vanishing possibility that the system reaches values corresponding to $K$ very far from $K_0$. But since the system has infinite time to reach these states it must be strictly impossible to do so in order to divide the phase space made up of all the angles in $\mathbb{Z}\alpha$.
	
	This would indeed be a novel kind of nonergodicity. It would involve regions of phase space of intermediate measure and different invariants (e.g. $K_0$) but which overlap each other in contravention of KAM theory. One way out of this, and to restore harmony with KAM theory where systems with different invariants stay in disjoint regions of phase space, is to ``promote'' $\theta_0$ to a phase variable, breaking the phase space up into disjoint strands, each corresponding to a different value of $\theta_0$. This is practical in the case where $\alpha \notin \Q$ and one has $K$,\footnote{Since $K_0 = \left< K \right>$, as we show later, knowledge of $K_0$ comes for free with $K$.}. In this case, one also has access to the unique value of $\theta_0$. The only way for $\Tilde{\theta} = K'\alpha = K\alpha$ for $K'\ne K$ is for there to be some non-zero integer $q : K\alpha \equiv (q + K)\alpha \mod{2\pi} \implies q\alpha \equiv 0 \mod{2\pi}$. But this implies that $\alpha = \frac{p}{q}2\pi \in \mathbb{Q}\pi$ for $p,q\in \mathbb{Z}$, in contradiction with $\alpha \notin \mathbb{Q}\pi$. Therefore $\theta_0 = \Tilde{\theta} - K\alpha$ is always a unique (up to $\varphi$) and available observable that no time will wash out. This is equally applicable to the case where $K$ runs off to infinity, its set has full measure, and the system \emph{is} ergodic. But if initial conditions affect the system, even if they are unknown, the system is nonergodic, but where initial conditions can be ignored, they ought to be, is part of the foundation of ergodic theory\cite{vonPlato1991}. If it is the case that there are different, finite ranges of $K$ that are reachable from different initial conditions then this trick of promoting intial conditions serves to prevent regions of overlap and harmonize the system with KAM theory. If there are no such finite ranges dependent on initial conditions, then those conditions can be forgotten, as the system is ergodic. 
	
	When examined carefully, their data do not necessitate that the system be nonergodic, even in the irrational case. To start with, Wang et al. show and cite theory that the system \emph{is} ergodic in the case that $\alpha$ is near-rational, such as when $\alpha$ is Liouville's number multiplied by $\pi$. This means it cannot be a simple distinction between the rational and irrational cases due to, for example, the availability of the initial conditions. A decisive proof that they are all ergodic is beyond the scope of this work, billiards in irrational triangles being a very open subject in dynamical systems\cite{gutkin2012billiard}, but there are more difficulties if it is nonergodic only in some ``strongly'' irrational regime.
	
    If the logarithmic growth discussed at the end of the last section is in fact the behavior of $n^*_K$, then the localization of $K$ around $K_0$ by an exponential when averaged over $\theta_0$ \cite[Fig.~3]{wang2014nonergodicity} does not speak to nonergodicity. It shows that $K$ remains exponentially close to its initial value whether looked at over time (the $n^*_K$ behavior as shown in \cite[Fig.~2]{wang2014nonergodicity}) or over initial conditions (as in \cite[Fig.~3]{wang2014nonergodicity}) which, taking $K$ as the observable, is what is expected in an ergodic system. 
    
    That this will always happen can be seen from $K$ itself. It only ever changes by one. Therefore it is, as the authors say, dense in its values up to $K_{\max}$, even if $\Km \rightarrow \infty$. It is also true that a leg collision is never more than one hit away from flipping the sign of $K$. This has the effect of distributing $K$ symmetrically about $K_0$; which ensures that the time average of $K$ \emph{is} $K_0$ as suggested by the numerical evidence. The average of $K$ over $\theta_0$ is $K_0$, almost by definition, as the choice of $K_0$ is a gauge freedom and independent of $\theta_0$. Thus the averages of $K$ will always converge to $K_0$.
	
	The fact that it is localized is of no consequence because while $K$ measures the size of subsets of the phase space, it does not measure the distances between velocity angles so remaining local near $K_0$ does not mean remaining near $\theta_0$. Even two angles whose $K$ numbers differ by one, and placed in the same quadrant by an appropriate $\varphi$, will be $\alpha \mod{\pi/2}$ apart. When we consider just the $K\alpha$ term in $\theta$, we are considering the circle folded up into one of its quadrants so $K\alpha$ must be taken $\mod{\pi/2}$.
	
	The largest hurdle for this new kind of nonergodicity is the fact that, if there is some range in $K$ that is unreachable, there must be a $\Km$ where one can guarantee that a double-leg hit does not follow\footnote{A leg hit of some kind must follow as $K$ can only reach $\Km$ via a collision with the hypotenuse.}. Since this depends on $\theta$, not just $K\alpha$, there must be some mechanism to ensure that $\varphi$ takes on the right value to prevent a double leg hit. This whole analysis relies on $\varphi$ being almost wholly decoupled from $K\alpha$ and a mechanism to couple them is not at all clear. Unless their is such a mechanism $K$ is unbounded, has full measure, and the system is ergodic.

\subsection{Concluding Remarks}
\label{sec:colnclusion}

    We have shown that the results presented in \cite{wang2014nonergodicity} are, in fact, consistent with the billiard path in a strongly irrational triangle being ergodic in its velocity angle. While we have failed to show the contrary is impossible, we have mounted an argument that the more likely case is that they are ergodic. We have also supplied a trick to re-harmonize the system with KAM theory if it is nonergodic in some regime. If this kind of nonergodicity is to be studied further, it  would be necissary to find a cross-over condition or behavior. To disprove it for this system, one could show that $K$ is unbounded when $\alpha \notin \mathbb{Q}\pi$, even for $\alpha = \phi$.
    

	Since this invariant measure has been fruitful in the study of billiards in irrational triangles, it may be interesting to see what similar measures can be constucted in the case of other irrational polygons where many questions remain open about periodic orbits\cite{gutkin2012billiard}. It is especially tempting to look for ones that take advantage of the substantial body of mathematical work on irrational rotations\cite{huveneers2009subdiffusive,Knill_2011,kim2003waiting}.
    
    What we regrettably have not touched on here is the cross-over in behavior from the rational case --- where closed orbits are dense in the initial velocity angle\cite{masur1986closed} and the flows are ergodic\cite{Vorobets_1996} --- to the irrational case where only a handful of closed trajectory types are known\cite{galperin2003periodic,schwartz2009obtuse}. Current theory is consistent with the system being ergodic in both cases: with a finite set of angles and unbounded $K$ increasing in increments of the orbit length in the rational case, and an infinite set of angles with a bijection to an unbounded $K$ in the irrational sense. These cases can coexist side-by-side as the closed orbits of irrational triangles are unstable\cite{hooper2007periodic} and therefore any change in $\alpha$ is enough to destroy a periodic orbit found for a particular $\theta_0$. 
    
    The author would like to thank Maxim Olshanii for introducing him to the problem as well as Aur\'elien Perrin and H\'el\`ene Perrin (no relation) for their support and patience while I worked on a problem, at best, tenuously related to my thesis.
    
\bibliography{Billiards.bib}

\begin{thebibliography}{16}%
\makeatletter
\providecommand \@ifxundefined [1]{%
 \@ifx{#1\undefined}
}%
\providecommand \@ifnum [1]{%
 \ifnum #1\expandafter \@firstoftwo
 \else \expandafter \@secondoftwo
 \fi
}%
\providecommand \@ifx [1]{%
 \ifx #1\expandafter \@firstoftwo
 \else \expandafter \@secondoftwo
 \fi
}%
\providecommand \natexlab [1]{#1}%
\providecommand \enquote  [1]{``#1''}%
\providecommand \bibnamefont  [1]{#1}%
\providecommand \bibfnamefont [1]{#1}%
\providecommand \citenamefont [1]{#1}%
\providecommand \href@noop [0]{\@secondoftwo}%
\providecommand \href [0]{\begingroup \@sanitize@url \@href}%
\providecommand \@href[1]{\@@startlink{#1}\@@href}%
\providecommand \@@href[1]{\endgroup#1\@@endlink}%
\providecommand \@sanitize@url [0]{\catcode `\\12\catcode `\$12\catcode
  `\&12\catcode `\#12\catcode `\^12\catcode `\_12\catcode `\%12\relax}%
\providecommand \@@startlink[1]{}%
\providecommand \@@endlink[0]{}%
\providecommand \url  [0]{\begingroup\@sanitize@url \@url }%
\providecommand \@url [1]{\endgroup\@href {#1}{\urlprefix }}%
\providecommand \urlprefix  [0]{URL }%
\providecommand \Eprint [0]{\href }%
\providecommand \doibase [0]{https://doi.org/}%
\providecommand \selectlanguage [0]{\@gobble}%
\providecommand \bibinfo  [0]{\@secondoftwo}%
\providecommand \bibfield  [0]{\@secondoftwo}%
\providecommand \translation [1]{[#1]}%
\providecommand \BibitemOpen [0]{}%
\providecommand \bibitemStop [0]{}%
\providecommand \bibitemNoStop [0]{.\EOS\space}%
\providecommand \EOS [0]{\spacefactor3000\relax}%
\providecommand \BibitemShut  [1]{\csname bibitem#1\endcsname}%
\let\auto@bib@innerbib\@empty
\bibitem [{\citenamefont {Wang}\ \emph {et~al.}(2014)\citenamefont {Wang},
  \citenamefont {Casati},\ and\ \citenamefont
  {Prosen}}]{wang2014nonergodicity}%
  \BibitemOpen
  \bibfield  {author} {\bibinfo {author} {\bibfnamefont {J.}~\bibnamefont
  {Wang}}, \bibinfo {author} {\bibfnamefont {G.}~\bibnamefont {Casati}},\ and\
  \bibinfo {author} {\bibfnamefont {T.}~\bibnamefont {Prosen}},\ }\bibfield
  {title} {\bibinfo {title} {Nonergodicity and localization of invariant
  measure for two colliding masses},\ }\href@noop {} {\bibfield  {journal}
  {\bibinfo  {journal} {Physical Review E}\ }\textbf {\bibinfo {volume} {89}},\
  \bibinfo {pages} {042918} (\bibinfo {year} {2014})}\BibitemShut {NoStop}%
\bibitem [{\citenamefont {Vorobets}(1996)}]{Vorobets_1996}%
  \BibitemOpen
  \bibfield  {author} {\bibinfo {author} {\bibfnamefont {Y.~B.}\ \bibnamefont
  {Vorobets}},\ }\bibfield  {title} {\bibinfo {title} {Ergodicity of billiards
  in polygons: explicit examples},\ }\href
  {https://doi.org/10.1070/rm1996v051n04abeh002989} {\bibfield  {journal}
  {\bibinfo  {journal} {Russian Mathematical Surveys}\ }\textbf {\bibinfo
  {volume} {51}},\ \bibinfo {pages} {756} (\bibinfo {year} {1996})}\BibitemShut
  {NoStop}%
\bibitem [{\citenamefont {Lichtenberg}\ and\ \citenamefont
  {Lieberman}(1983)}]{lichtenberg1983regular}%
  \BibitemOpen
  \bibfield  {author} {\bibinfo {author} {\bibfnamefont {A.}~\bibnamefont
  {Lichtenberg}}\ and\ \bibinfo {author} {\bibfnamefont {M.}~\bibnamefont
  {Lieberman}},\ }\href@noop {} {\emph {\bibinfo {title} {Regular and
  Stochastic Motion}}},\ Vol.\ \bibinfo {volume} {277}\ (\bibinfo  {publisher}
  {Springer-Verlag New York},\ \bibinfo {year} {1983})\BibitemShut {NoStop}%
\bibitem [{\citenamefont {Cipra}\ \emph {et~al.}(1995)\citenamefont {Cipra},
  \citenamefont {Hanson},\ and\ \citenamefont {Kolan}}]{cipra1995periodic}%
  \BibitemOpen
  \bibfield  {author} {\bibinfo {author} {\bibfnamefont {B.}~\bibnamefont
  {Cipra}}, \bibinfo {author} {\bibfnamefont {R.~M.}\ \bibnamefont {Hanson}},\
  and\ \bibinfo {author} {\bibfnamefont {A.}~\bibnamefont {Kolan}},\ }\bibfield
   {title} {\bibinfo {title} {Periodic trajectories in right-triangle
  billiards},\ }\href {https://doi.org/10.1103/PhysRevE.52.2066} {\bibfield
  {journal} {\bibinfo  {journal} {Phys. Rev. E}\ }\textbf {\bibinfo {volume}
  {52}},\ \bibinfo {pages} {2066} (\bibinfo {year} {1995})}\BibitemShut
  {NoStop}%
\bibitem [{Note1()}]{Note1}%
  \BibitemOpen
  \bibinfo {note} {This occurs at $\alpha /2 = \protect \nicefrac {\pi }{4}$
  and gives $\left <p\right >^{-1} = 2.7741...$ a difference of 2\%. This also
  constitutes a maximum for $\left <p\right >$}\BibitemShut {NoStop}%
\bibitem [{Note2()}]{Note2}%
  \BibitemOpen
  \bibinfo {note} {Since $K_0 = \left < K \right >$, as we show later,
  knowledge of $K_0$ comes for free with $K$.}\BibitemShut {Stop}%
\bibitem [{\citenamefont {von Plato}(1991)}]{vonPlato1991}%
  \BibitemOpen
  \bibfield  {author} {\bibinfo {author} {\bibfnamefont {J.}~\bibnamefont {von
  Plato}},\ }\bibfield  {title} {\bibinfo {title} {Boltzmann's ergodic
  hypothesis},\ }\href {https://doi.org/10.1007/BF00384333} {\bibfield
  {journal} {\bibinfo  {journal} {Archive for History of Exact Sciences}\
  }\textbf {\bibinfo {volume} {42}},\ \bibinfo {pages} {71} (\bibinfo {year}
  {1991})}\BibitemShut {NoStop}%
\bibitem [{\citenamefont {Gutkin}(2012)}]{gutkin2012billiard}%
  \BibitemOpen
  \bibfield  {author} {\bibinfo {author} {\bibfnamefont {E.}~\bibnamefont
  {Gutkin}},\ }\bibfield  {title} {\bibinfo {title} {Billiard dynamics: An
  updated survey with the emphasis on open problems},\ }\href@noop {}
  {\bibfield  {journal} {\bibinfo  {journal} {Chaos: An Interdisciplinary
  Journal of Nonlinear Science}\ }\textbf {\bibinfo {volume} {22}},\ \bibinfo
  {pages} {026116} (\bibinfo {year} {2012})}\BibitemShut {NoStop}%
\bibitem [{Note3()}]{Note3}%
  \BibitemOpen
  \bibinfo {note} {A leg hit of some kind must follow as $K$ can only reach
  $K_{\protect \qopname \relax m{max}}$ via a collision with the
  hypotenuse.}\BibitemShut {Stop}%
\bibitem [{\citenamefont {Huveneers}(2009)}]{huveneers2009subdiffusive}%
  \BibitemOpen
  \bibfield  {author} {\bibinfo {author} {\bibfnamefont {F.}~\bibnamefont
  {Huveneers}},\ }\bibfield  {title} {\bibinfo {title} {Subdiffusive behavior
  generated by irrational rotations},\ }\href@noop {} {\bibfield  {journal}
  {\bibinfo  {journal} {Ergodic Theory and Dynamical Systems}\ }\textbf
  {\bibinfo {volume} {29}},\ \bibinfo {pages} {1217} (\bibinfo {year}
  {2009})}\BibitemShut {NoStop}%
\bibitem [{\citenamefont {Knill}\ and\ \citenamefont
  {Tangerman}(2011)}]{Knill_2011}%
  \BibitemOpen
  \bibfield  {author} {\bibinfo {author} {\bibfnamefont {O.}~\bibnamefont
  {Knill}}\ and\ \bibinfo {author} {\bibfnamefont {F.}~\bibnamefont
  {Tangerman}},\ }\bibfield  {title} {\bibinfo {title} {Self-similarity and
  growth in birkhoff sums for the golden rotation},\ }\href
  {https://doi.org/10.1088/0951-7715/24/11/006} {\bibfield  {journal} {\bibinfo
   {journal} {Nonlinearity}\ }\textbf {\bibinfo {volume} {24}},\ \bibinfo
  {pages} {3115} (\bibinfo {year} {2011})}\BibitemShut {NoStop}%
\bibitem [{\citenamefont {Kim}\ and\ \citenamefont
  {Seo}(2003)}]{kim2003waiting}%
  \BibitemOpen
  \bibfield  {author} {\bibinfo {author} {\bibfnamefont {D.~H.}\ \bibnamefont
  {Kim}}\ and\ \bibinfo {author} {\bibfnamefont {B.~K.}\ \bibnamefont {Seo}},\
  }\bibfield  {title} {\bibinfo {title} {The waiting time for irrational
  rotations},\ }\href@noop {} {\bibfield  {journal} {\bibinfo  {journal}
  {Nonlinearity}\ }\textbf {\bibinfo {volume} {16}},\ \bibinfo {pages} {1861}
  (\bibinfo {year} {2003})}\BibitemShut {NoStop}%
\bibitem [{\citenamefont {Masur}\ \emph {et~al.}(1986)\citenamefont {Masur}
  \emph {et~al.}}]{masur1986closed}%
  \BibitemOpen
  \bibfield  {author} {\bibinfo {author} {\bibfnamefont {H.}~\bibnamefont
  {Masur}} \emph {et~al.},\ }\bibfield  {title} {\bibinfo {title} {Closed
  trajectories for quadratic differentials with an application to billiards},\
  }\href@noop {} {\bibfield  {journal} {\bibinfo  {journal} {Duke mathematical
  journal}\ }\textbf {\bibinfo {volume} {53}},\ \bibinfo {pages} {307}
  (\bibinfo {year} {1986})}\BibitemShut {NoStop}%
\bibitem [{\citenamefont {Galperin}\ and\ \citenamefont
  {Zvonkine}(2003)}]{galperin2003periodic}%
  \BibitemOpen
  \bibfield  {author} {\bibinfo {author} {\bibfnamefont {G.}~\bibnamefont
  {Galperin}}\ and\ \bibinfo {author} {\bibfnamefont {D.}~\bibnamefont
  {Zvonkine}},\ }\bibfield  {title} {\bibinfo {title} {Periodic billiard
  trajectories in right triangles and right-angled tetrahedra},\ }\href@noop {}
  {\bibfield  {journal} {\bibinfo  {journal} {Regul. Chaotic Dyn}\ }\textbf
  {\bibinfo {volume} {8}},\ \bibinfo {pages} {29} (\bibinfo {year}
  {2003})}\BibitemShut {NoStop}%
\bibitem [{\citenamefont {Schwartz}(2009)}]{schwartz2009obtuse}%
  \BibitemOpen
  \bibfield  {author} {\bibinfo {author} {\bibfnamefont {R.~E.}\ \bibnamefont
  {Schwartz}},\ }\bibfield  {title} {\bibinfo {title} {Obtuse triangular
  billiards ii: One hundred degrees worth of periodic trajectories},\
  }\href@noop {} {\bibfield  {journal} {\bibinfo  {journal} {Experimental
  Mathematics}\ }\textbf {\bibinfo {volume} {18}},\ \bibinfo {pages} {137}
  (\bibinfo {year} {2009})}\BibitemShut {NoStop}%
\bibitem [{\citenamefont {Hooper}(2007)}]{hooper2007periodic}%
  \BibitemOpen
  \bibfield  {author} {\bibinfo {author} {\bibfnamefont {W.~P.}\ \bibnamefont
  {Hooper}},\ }\bibfield  {title} {\bibinfo {title} {Periodic billiard paths in
  right triangles are unstable},\ }\href@noop {} {\bibfield  {journal}
  {\bibinfo  {journal} {Geometriae Dedicata}\ }\textbf {\bibinfo {volume}
  {125}},\ \bibinfo {pages} {39} (\bibinfo {year} {2007})}\BibitemShut
  {NoStop}%
\end{thebibliography}%

\end{document}